\documentstyle[preprint,aps]{revtex}
\tighten
\draft
\widetext
\preprint{HUTP-98/A090, NUB 3193}
\bigskip
\bigskip
\begin{document}
\title{Flavor ``Conservation'' and Hierarchy in TeV-scale
Supersymmetric Standard Model}
\medskip

\author{Zurab Kakushadze\footnote{E-mail: 
zurab@string.harvard.edu}}

\bigskip
\address{Lyman Laboratory of Physics, Harvard University,
Cambridge,  MA 02138\\
and\\
Department of Physics, Northeastern University, Boston, MA 02115}

\date{February 11, 1999}
\bigskip
\medskip
\maketitle

\begin{abstract}
{}Recently a TeV-scale Supersymmetric Standard Model (TSSM) was proposed
in which the gauge coupling unification is as precise (at one loop) as 
in the MSSM, and occurs in the TeV range. Proton stability in the TSSM
is due to an anomaly free ${\bf Z}_3\otimes {\bf Z}_3$ discrete gauge 
symmetry, which is also essential for successfully generating neutrino 
masses in the desirable range. In this paper we show that the TSSM admits 
anomaly free non-Abelian {\em discrete} flavor gauge symmetries (based on a
left-right product 
tetrahedral group) which together with a ``vector-like'' Abelian (discrete)
flavor gauge symmetry suppresses dangerous higher dimensional operators 
corresponding to flavor
changing neutral currents (FCNCs) to an acceptable level. Discrete
flavor gauge 
symmetries are more advantageous compared with continuous flavor gauge
symmetries as the latter must be broken, which generically results in
unacceptably large {\em gauge} mediated flavor violation. In contrast, in
the case of discrete flavor gauge symmetries the only possibly dangerous 
sources of flavor violation
either come from the corresponding ``bulk'' flavon (that is, flavor symmetry
breaking Higgs) exchanges, or are induced by flavon VEVs. These sources of 
flavor violation, however, are adequately
suppressed by the above flavor gauge symmetries for the string scale $\sim
10-100~{\mbox{TeV}}$.    
\end{abstract}

\section{Introduction}

{}Recently it has become clear that the discovery of D-branes \cite{polchi}
may have profound phenomenological 
implications\footnote{For recent developments, see, {\em e.g.}, 
\cite{witt,lyk,TeV,dien,3gen,anto,ST,3gen1,TeVphen,BW}.}. In particular, 
the Standard Model gauge fields (as well as the corresponding charged
matter) may reside inside of $p\leq 9$ spatial dimensional $p$-branes (or a
set of overlapping branes), while gravity lives in a larger (10 or 11)
dimensional bulk of space-time. This ``Brane World'' scenario appears to be
flexible enough so that satisfying various requirements such as gauge and
gravitational coupling unification, dilaton stabilization and weakness of
the Standard Model gauge couplings seems to be possible \cite{BW} within
this framework (provided that the Standard Model fields live on branes 
with $3<p<9$). This suggests that the brane world scenario might be 
a coherent picture for describing our universe \cite{BW}\footnote{The brane
world picture in the effective field theory context was discussed in
\cite{early,shif}.}.

{}In string theory, which is the only known theory that consistently
incorporates quantum gravity, the gauge and gravitational couplings are 
expected to unify (up to an order one factor due to various thresholds 
\cite{kap,BF}) at the string scale
$M_s=1/\sqrt{\alpha^\prime}$. In the brane world scenario the string scale
is {\em a priori} undetermined, and can be anywhere between the electroweak 
scale $M_{ew}$ and the Planck scale $M_P=1/\sqrt{G_N}$ (where $G_N$ is
the Newton's constant). Thus, if we assume that the bulk is ten
dimensional, then the four dimensional gauge and gravitational couplings 
scale as\footnote{For illustrative purposes here we are using the 
corresponding tree-level relations in Type I (or Type I$^\prime$) theory.} 
$\alpha\sim g_s/V_{p-3} M_s^{p-3}$ respectively 
$G_N\sim g_s^2/V_{p-3} V_{9-p} M_s^8$, where $g_s$ is the string coupling, 
and $V_{p-3}$ and $V_{9-p}$ are the compactification volumes inside and 
transverse to the $p$-branes, respectively. 
For $3<p<9$ there are two {\em a priori} independent volume factors, and, 
for the fixed gauge coupling $\alpha$ (at the
unification, that is, string scale) and four dimensional Planck scale 
$M_P$, the string scale is not determined. Based on this fact, in
\cite{witt} it was proposed that the gauge and gravitational coupling 
unification problem\footnote{For a review of the gauge and gravitational 
coupling unification problem in the perturbative heterotic string context, 
see, {\em e.g.}, \cite{Dienes}, and references therein. In the Type I
context the discussions on this issue can be found in \cite{CKM,BW}.} can 
be ameliorated in this context by lowering the string scale $M_s$ down to 
the GUT scale $M_{\small{GUT}}\approx 2\times 10^{16}~{\mbox{GeV}}$
\cite{gut}\footnote{By the GUT scale here we mean the usual scale of gauge 
coupling unification in the MSSM obtained by extrapolating the LEP data in 
the assumption of the standard ``desert'' scenario.}. In \cite{lyk} it was 
noticed that $M_s$ can be lowered all the way down to TeV.

{}More recently it was proposed in \cite{TeV} that $M_s$ 
as well as\footnote{Note that the string scale $M_s$ cannot be too much
lower than the fundamental Planck scale or else the string coupling $g_s$
as well as all the gauge couplings would come out too small contradicting 
the experimental data.} the fundamental (10 or 11 dimensional) Planck scale 
can be around TeV. The observed weakness of the four dimensional 
gravitational coupling then requires the presence of at least two large 
($\gg 1/M_s$) compact directions transverse to the $p$-branes on which the 
Standard Model fields are localized. A general discussion of possible brane 
world embeddings of such a scenario was given in \cite{anto,ST,BW}. 
In \cite{TeVphen} various non-trivial phenomenological 
issues were discussed in the context of the TeV-scale brane world scenario, 
and it was argued that this possibility does not appear to be automatically 
ruled out\footnote{For other recent 
works on TeV-scale string/gravity
scenarios, see, {\em e.g.}, 
\cite{dien,flavor1,flavor,neutrino,zura,AB,zura1,henry,Pomarol,related}. 
For other scenarios with
lowered string scale and related works, see, {\em e.g.},
\cite{other}. TeV-scale compactifications were studied in \cite{quiros} in
the context of supersymmetry breaking.}. 

{}In such a scenario, however, as well as in any scenario with $M_s\ll
M_{\small{GUT}}$, the gauge coupling unification at $M_s$ would have 
to arise via a mechanism rather different from the usual MSSM 
unification which occurs with a remarkable precision 
\cite{gut}. In the brane world picture there appears to exist such a 
mechanism \cite{dien} for lowering the unification scale. 
Thus, let the ``size'' $R$ of the compact dimensions inside of the 
$p$-brane (where $p>3$) be somewhat large compared with $1/M_s$. Then the 
evolution of the gauge couplings above the Kaluza-Klein (KK) 
threshold $1/R$ is no longer logarithmic but power-like \cite{TV}. 
This observation was used in
\cite{dien} to argue that the gauge coupling unification might occur at a
scale (which in the brane world context would be identified with the string
scale) much lower than $M_{\small{GUT}}$. For successfully implementing
this mechanism, however, it is also necessary to find a concrete extension 
of the MSSM such that the unification prediction is just as precise as in 
the MSSM (at least at one loop). In fact, one could also require that such 
an extension {\em explain} why couplings unify in the MSSM at all, that is, 
why the unification in the MSSM is {\em not} just an ``accident'' (assuming 
that the TeV-scale brane world scenario has the pretense of replacing the 
old framework).

{}In \cite{zura} a TeV-scale Supersymmetric Standard Model (TSSM) was proposed
in which the gauge coupling unification indeed occurs via such a higher 
dimensional mechanism. Moreover, the unification in the TSSM is as precise 
(at one loop) as in the MSSM, and occurs in the TeV 
range\footnote{By the TeV range we do {\em not} necessarily mean that 
$M_s\sim 1~{\mbox{TeV}}$. In fact, as was argued in \cite{zura}, 
the gauge coupling unification constraints seem to imply that $M_s$ cannot 
really be lower than $10-100~{\mbox{TeV}}$.}. 
In particular, the key ingredient of the TSSM is the presence
of new (compared with the MSSM) light states neutral under $SU(3)_c\otimes
SU(2)_w$ but charged under $U(1)_Y$ whose mass scale is around that of the
electroweak Higgs doublets. It is the heavy Kaluza-Klein tower
corresponding to these new states which makes it possible to satisfy the
requirement that the unification in the TSSM be as
precise (at one loop) as in the MSSM. In fact, as was pointed out in 
\cite{zura}, after a rather systematic 
search the TSSM was the only (simple) solution found for this constraint. 
The TSSM also explains why the unification in the MSSM is not an accident -
if the TSSM is indeed (a part of) the correct description of nature above
the electroweak scale, then the gauge coupling unification in the MSSM is
explained by the present lack of experimental data which leads to the
standard ``desert'' assumption. 

{}One of the most obvious worries with the TeV-scale brane world scenario
in general is the proton stability problem - higher dimensional baryon and
lepton number violating operators are generically suppressed only by powers
of $1/M_s$ with $M_s$ in the TeV range, which is inadequate to ensure the
observed proton longevity. In \cite{zura1} it was shown that
introduction of the new states responsible for the gauge coupling
unification in the TSSM also allows to gauge anomaly free discrete symmetries
which suppress dangerous higher dimensional operators and stabilize
proton. In particular, in \cite{zura1} an anomaly free   
${\bf Z}_3\otimes {\bf Z}_3$ discrete gauge symmetry which makes proton 
completely stable was explicitly constructed. In \cite{zura1} it was also
argued that this discrete gauge symmetry is essential for successfully
generating neutrino masses in the desirable range via a higher dimensional
mechanism recently proposed in \cite{neutrino}. In particular, in
\cite{zura1} it was pointed out that certain dimension 5 lepton number
violating operators must be
absent or else unacceptably large Majorana neutrino masses would be
generated upon the electroweak symmetry breaking. 
More concretely, from this viewpoint unsuppressed dimension 5 operators of
the form $LLH_+H_+$ would be disastrous, where $L$ is the $SU(2)_w$ doublet
containing a left-handed neutrino and the corresponding charged lepton (we
are suppressing the flavor indices), and $H_+$ is the electroweak Higgs
doublet with the hypercharge $+1$.  
The ${\bf Z}_3\otimes {\bf Z}_3$ discrete gauge
symmetry mentioned above stabilizes proton and forbids these dimension 
5 operators in one shot.    

{}The fact that the gauge coupling unification, proton stability and
neutrino mass problems can be solved within the TSSM suggests that 
it is reasonable to take the TSSM as a starting point for 
addressing many other open questions that the TeV-scale brane world
scenario faces\footnote{Supersymmetry breaking in the TSSM was recently
discussed in \cite{Pomarol} in the context of the Scherk-Schwarz mechanism 
\cite{SS}.}. In this paper we will focus on another obvious worry with
the TeV-scale brane world scenario: flavor changing neutral currents
(FCNCs). Here we will take the approach where we will try
to find possible solutions to this problem in ways consistent with other
features described above such as proton stability and neutrino masses.   
This appears to imply that the number of viable possibilities is rather 
limited which allows to explore them in a systematic fashion.

{}Certain generic aspects of the FCNC problem in the context of the 
TeV-scale 
brane world scenario were discussed in \cite{flavor1,flavor}. Thus, in
\cite{flavor1} global flavor symmetries were considered in the context of
generating the 
desired flavor hierarchy. It was pointed out in \cite{flavor1}
that hierarchical breaking of such flavor symmetries on ``distant'' branes
could account for the observed fermion mass hierarchy in the Standard
Model. Breaking global flavor symmetries on ``distant'' branes would be 
compatible with the current experimental bounds on FCNCs as the only
possibly dangerous sources of flavor violation induced by such breaking come 
from the corresponding ``bulk'' flavon (that is, flavor symmetry breaking
Higgs) exchanges. The latter are adequately suppressed by the volume
of the corresponding large dimensions (required to be present in the
TeV-scale brane world scenario) transverse to the $p$-branes on which the
Standard Model fields are localized. However, global continuous symmetries 
or non-gauge discrete symmetries may not be completely adequate in this
context. First, generically quantum gravity effects (wormholes, {\em etc.}) 
are expected to violate such global symmetries and induce effective 
flavor violating higher dimensional operators which would only be suppressed by
the corresponding powers of $1/M_s$ \cite{gil}. Second, it is believed that 
there are no global symmetries in string theory 
\cite{global}\footnote{These two observations may not be completely 
unrelated.}. This then implies that we should consider either continuous or 
discrete flavor {\em gauge} symmetries. 
Discrete gauge symmetries are believed to be stable under quantum gravity 
effects \cite{Krauss} (also see, {\em e.g.}, \cite{Wilczek}), and they do 
arise in string theory. At first it might appear that continuous flavor
gauge symmetries would suppress FCNCs much more ``efficiently'' than 
discrete flavor gauge symmetries. However, it was pointed out in
\cite{flavor} that gauging continuous flavor symmetries immediately runs
into the following problem. First, it was shown in \cite{flavor} that
Abelian continuous flavor gauge symmetries cannot do the job of adequately
suppressing all dangerous FCNCs (for instance, those relevant for the Kaon
system). This implies that we would have to consider {\em
non-Abelian} continuous flavor gauge symmetries. However, these gauge
symmetries must be broken at some scale (below $M_s$) or else the 
corresponding massless gauge bosons would give rise to experimentally 
excluded long range forces. Also, the observed fermion hierarchy in the
Standard Model is incompatible with such unbroken flavor symmetries.
In \cite{flavor} it was shown that even if
these flavor symmetries are ``bulk'' gauge symmetries\footnote{Note that
such flavor gauge symmetries would actually have to be ``bulk'' gauge
symmetries. Thus, if the corresponding gauge bosons are localized on the
same $p$-branes as the Standard Model fields, the tree-level exchanges of
the horizontal gauge boson(s) would ultimately reintroduce the exact same
types of FCNCs (with unacceptable strengths)
as those which the flavor gauge symmetry was supposed to
suppress in the first place.}, the gauge mediated flavor violation
triggered by the flavor symmetry breaking generically is not so small
after all, and can lead to unacceptably large FCNCs. More precisely, it was
shown in \cite{flavor} that the flavor mediated FCNCs are not at all 
suppressed in the case of 2 large dimensions transverse to the $p$-branes, 
and an adequate suppression can only be achieved if the number of large
transverse dimensions is $\geq 4$ (provided that $M_s$ is not lower than
$10-100~{\mbox{TeV}}$). {\em A priori} this fact may not look like a big
deal, albeit it would rule out (provided that there are no solutions
to the problem other than gauging continuous flavor symmetries in the bulk)
the cases with less than 4 large transverse dimensions. Here it is reasonable
to ask whether one can find a much more ``efficient'' way of solving the
FCNC problem in the TeV-scale brane world context. This becomes especially
desirable taking into account that, as was pointed out in \cite{zura1}, 
in the brane world (that is, string
theory) framework it might be necessary to have 2 and only 2 large
transverse directions (at least in the TSSM context)
if one would like the higher dimensional mechanism of
\cite{neutrino} for generating the correct neutrino masses to work (and, at
the same time, be compatible with ${\cal N}=1$ supersymmetry which seems to
be essential for the gauge coupling unification \cite{zura}). Moreover, as
was pointed out in \cite{zura1}, the discussion of dilaton stabilization in
\cite{BW}, which takes into account various observations of \cite{BaDi} as
well as the 
explicit mechanisms of dilaton stabilization \cite{kras}, suggests
that to achieve the latter we do not seem to be allowed to have
more\footnote{Note that having only one large transverse dimension is
experimentally excluded \cite{TeV} for otherwise there would be deviations
from the Newtonian gravity over solar system distances.} than
2 large transverse dimensions in the TeV-scale brane world context. The 
recent discussion in \cite{AB} also suggests that radius stabilization at
large values in the brane world framework seems to favor having 2 large
transverse directions. These considerations all indicate that the case of 2
large transverse directions could be the most interesting one, so letting
it be ruled out so easily may not be desirable.    
 
{}Motivated by the above considerations, in this paper we consider
anomaly free non-Abelian {\em discrete} flavor gauge symmetries for suppressing
FCNCs in the TSSM. The advantage of gauging discrete rather than continuous
flavor symmetries is that, unlike in the latter case, 
upon breaking the former no gauge mediated flavor
violation occurs whatsoever - there are no horizontal gauge bosons to start
with. Thus, just as in the case of global flavor symmetries, the only
possibly dangerous sources of flavor violation come from the corresponding
``bulk'' flavon exchanges which are adequately suppressed. On the other
hand, as we have already pointed out, considering gauge rather than global
symmetries appears to be 
necessary in this context since we are dealing with a theory
where quantum gravity becomes strongly coupled at energies around $M_s$
which sets the ``cut-off'' for the induced effective flavor violating
higher dimensional operators. Gauging non-Abelian discrete symmetries in
the TSSM is not completely trivial - we must ensure that these gauge
symmetries are anomaly free. This puts tight constraints on possible flavor
symmetries we can gauge, especially that they must be compatible with other
anomaly free gauge symmetries (such as the ${\bf Z}_3\otimes {\bf Z}_3$
discrete gauge symmetry responsible for proton longevity in the TSSM). In
fact, the number of possibilities we find is rather limited, and the {\em
conclusive} solution to the FCNC problem we present in this paper is based
on a left-right product (non-Abelian) 
tetrahedral group $T_L\otimes T_R$ (which
can be viewed as a discrete subgroup of the $SU(2)_L\otimes SU(2)_R$ flavor
symmetry group) accompanied by a ``vector-like'' $U(1)_V$ flavor gauge
symmetry (or its appropriate discrete subgroup). The non-Abelian part of the
flavor group together with the $U(1)_V$ subgroup 
adequately suppresses FCNCs
(for any number $\geq 2$ of large transverse dimensions, and the string
scale $M_s\sim 10-100~{\mbox{TeV}}$)
provided that the flavor symmetry breaking Higgses (flavons) are ``bulk''
fields.

{}The rest of this paper is organized as follows. In section II we briefly
review the TSSM proposed in \cite{zura}. We mainly focus on the light
spectrum as the heavy KK modes are not going to be relevant for the
subsequent discussions. We also briefly review the discrete gauge
symmetries proposed in \cite{zura1} which ensure proton stability and
successful generation of small neutrino masses. These discrete gauge
symmetries are relevant in the following as the flavor symmetries we are
going to gauge must be anomaly free without spoiling the anomaly freedom
condition for the former discrete gauge symmetries. In section III we
briefly review the discussion in \cite{flavor}, which we tailor to our
purposes in this paper. In section IV we explicitly construct an 
anomaly free
discrete flavor gauge symmetry which, as we show, adequately suppresses
FCNCs in the TSSM. There we also discuss various phenomenological 
implications (including possible collider signatures)  
of these symmetries together with other symmetries in the TSSM. 
In section V we briefly summarize our results. 

{}We have been recently informed that some issues related to flavor 
violation in the TeV-scale brane world context are also going to be
discussed in \cite{AHS,BDN}. 

\section{The TSSM} 

{}In this section we briefly review the TSSM proposed in \cite{zura}. 
The gauge group of this model is the same as in the MSSM, that is, 
$SU(3)_c\otimes SU(2)_w \otimes U(1)_Y$. The light 
spectrum\footnote{By the light spectrum we mean the states which 
are massless before the supersymmetry/electroweak symmetry breaking.} 
of the model is ${\cal N}=1$ supersymmetric, and along with the vector 
superfields $V$ transforming in the adjoint
of $SU(3)_c\otimes SU(2)_w \otimes U(1)_Y$ we also have the following 
chiral superfields
(corresponding to the matter and Higgs particles):
\begin{eqnarray}
 && Q_i=3\times ({\bf 3},{\bf 2})(+1/3)~,~~~
 D_i=3\times ({\overline {\bf 3}},{\bf 1})(+2/3)~,~~~
 U_i=3\times ({\overline {\bf 3}},{\bf 1})(-4/3)~,\nonumber\\
 && L_i=3\times ({\bf 1},{\bf 2})(-1)~,~~~E_i=3\times ({\bf 1},
 {\bf 1})(+2)~,~~~
 N_i=3\times ({\bf 1},{\bf 1})(0)~,\nonumber\\
 && H_+=({\bf 1},{\bf 2})(+1)~,~~~H_-=({\bf 1},{\bf 2})(-1)~,\nonumber\\
 && F_+=({\bf 1},{\bf 1})(+2)~,~~~F_-=({\bf 1},{\bf 1})(-2)~.\nonumber
\end{eqnarray}
Here the $SU(3)_c\otimes SU(2)_w$ quantum numbers are given in bold font, 
whereas the $U(1)_Y$ hypercharge is given in parentheses. The three 
generations $(i=1,2,3)$ of quarks 
and leptons are given by $Q_i,D_i,U_i$ respectively $L_i,E_i,N_i$ 
(the chiral superfields $N_i$ correspond 
to the right-handed neutrinos), whereas $H_\pm$ correspond to the 
electroweak Higgs doublets. Note that the chiral superfields $F_\pm$ 
are {\em new}: they were not present in the MSSM. 

{}The massive spectrum of the TSSM contains Kaluza-Klein (KK) states. 
These states correspond to compact $p-3$ dimensions inside of the 
$p$-branes ($p=4$ or 5) on which the gauge fields are localized. 
The heavy KK levels are populated by ${\cal N}=2$ supermultiplets with 
the quantum numbers given by $({\widetilde V},{\widetilde H}_+,
{\widetilde H}_-,{\widetilde F}_+,{\widetilde F}_-)$, where 
${\widetilde V}$ stands for the ${\cal N}=2$ vector superfield transforming
in the adjoint of $SU(3)_c\otimes SU(2)_w\otimes U(1)_Y$, whereas 
${\widetilde H}_\pm,{\widetilde F}_\pm$ are the 
${\cal N}=2$ hypermultiplets
with the gauge quantum numbers of $H_\pm,F_\pm$.  
(The exact massive KK spectrum, which is not going to be important in the
subsequent discussions, can be found in \cite{zura,zura1}.)
Here we would like to point out some of the features of the model which 
are going to be relevant for discussions in this section as well as 
sections III and IV. Note that the massless superfields 
$V,H_\pm,F_\pm$ have heavy KK counterparts. We can think about 
these states together with the corresponding heavy KK modes as arising 
upon compactification of a $p+1$ dimensional theory on a $p-3$ dimensional 
compact space with the volume $V_{p-3}$. On the other hand, the 
massless superfields $Q_i,D_i,U_i,L_i,E_i,N_i$ do not possess heavy 
KK counterparts corresponding to these $p-3$ dimensions. 
(A concrete mechanism for localizing these fields in the brane world
context was discussed in detail in 
\cite{zura,zura1}.)

{}The gauge coupling unification in the TSSM is just as precise (at one
loop) as in the MSSM \cite{zura}, and the unification scale $M_s$ is in the
TeV range (provided that the mass scale of the superfields $F_\pm$ is around
that of the electroweak Higgs doublets). 
The lowering of the unification scale here occurs 
along the lines of \cite{dien} due to the power-like running of the gauge
couplings above the KK threshold scale \cite{TV}. On the other hand, the
fact that the one-loop unification in the TSSM is as precise as in the MSSM
crucially depends on the KK tower of the new states $F_\pm$. (The actual 
value of $M_s$ \cite{zura,zura1}
depends on the volume $V_{p-3}$ which is assumed to be (relatively) large, 
that is, $V_{p-3}/(2\pi)^{p-3}\gg 1/M_s^{p-3}$.) The unified gauge 
coupling $\alpha$ in the TSSM is small\footnote{Note, however,
that the true loop expansion 
parameter is of order one \cite{zura} for it 
is enhanced by a factor proportional to the number of the heavy 
KK states which is large. (This enhancement is analogous to that in large
$N$ gauge theories \cite{thooft}.) Nonetheless, as explained in \cite{zura},
one-loop corrections to the gauge couplings are still dominant due
to supersymmetry.}. Thus, for instance, for 
$M_s\simeq 10~{\mbox{TeV}}$ we have $\alpha\simeq 1/37.5$.  

\subsection{Proton Stability and Neutrino Masses}

{}To stabilize proton, in \cite{zura1} a ${\bf Z}_3\otimes {\bf Z}_3$ 
discrete gauge symmetry was introduced. Following \cite{zura1} we will
refer to this discrete gauge symmetry as ${\widetilde {\cal L}}_3\otimes  
{\widetilde {\cal R}}_3$. The corresponding discrete charges (which are
conserved modulo 3) are given by:
\begin{eqnarray}
 &&Q:~(0,0)~,~~~D:~(0,+1)~,~~~U:~(0,-1)~,\nonumber\\
 &&L:~(+1,0)~,~~~E:~(-1,+1)~,~~~N:~(-1,-1)~,\nonumber\\
 &&H_+:~(0,+1)~,~~~H_-:~(0,-1)~,\nonumber\\
 &&F_+:~(0,-1)~,~~~F_-:~(0,+1)~.\nonumber
\end{eqnarray}
In \cite{zura1} it was shown that this discrete gauge symmetry is anomaly
free. In particular, it satisfies the anomaly freedom conditions discussed
in \cite{IR1}. In fact, this gauge symmetry satisfies even stronger
conditions as it can be embedded into an anomaly free continuous gauge
symmetry which we refer to $U(1)_{\cal L}\otimes U(1)_{\cal R}$ \cite{zura1}.
This fact will be useful in section IV.

{}In \cite{zura1} it was shown that the ${\widetilde {\cal L}}_3\otimes  
{\widetilde {\cal R}}_3$ discrete gauge symmetry stabilizes proton. Thus,
it forbids all dangerous baryon and lepton number violating operators
potentially leading to proton decay. In particular, dimension 4
\cite{wein}, dimension 5 (such as $QQQL$) as well as all other higher
dimensional operators of this type are forbidden by this discrete gauge
symmetry.   

{}The ${\widetilde {\cal L}}_3\otimes  
{\widetilde {\cal R}}_3$ discrete gauge symmetry also forbids the dangerous
lepton number violating dimension 5 operator $LLH_+H_+$ which would be
disastrous\footnote{Note that this operator is precisely the one
responsible for generating the correct neutrino masses in the old
``see-saw'' mechanism \cite{see-saw} in scenarios with high 
$M_s\sim M_{\small{GUT}}$.} for neutrino masses \cite{zura1} - 
this operator is suppressed only by $1/M_s$ (with $M_s$ in the TeV range),
and would result in unacceptably large Majorana neutrino masses $m_\nu\sim
M_{ew}^2/M_s$ upon the electroweak symmetry breaking. 
The desirable Dirac neutrino masses in the TSSM can then be
generated via the higher dimensional mechanism of \cite{neutrino}.

\section{Flavor Violation in the TeV-scale Brane World Scenario}

{}In this section we would like to review some of the discussions of 
\cite{flavor} on flavor violation in the TeV-scale brane world
scenario. In particular, one of the important points in \cite{flavor} 
is the estimate for the
expected flavor violation coming from the ``bulk'' flavon as well as horizontal
gauge boson (in the cases with continuous flavor gauge symmetries)
exchanges. The latter are dominant, and lead to unacceptably large flavor
violation unless the number of large transverse dimensions is $\geq 4$.

\subsection{Why Do We Need Flavor Symmetries?}

{}There are tight experimental bounds on the flavor changing neutral
currents \cite{PDG}, the most constraining being the Kaon system. Thus, let
us consider the constraints from the mass splitting between $K^0_S$ and
$K^0_L$. The lowest dimensional operators relevant in this context are the
dimension 6 four-fermion operators of the form (for simplicity our
notations here are symbolic, and we suppress the corresponding color and
Lorentz indices as well as the chiral projection operators for the left-
and right-handed fields): 
\begin{equation}\label{kaon}
 \xi ({\overline s}d)^2~.
\end{equation}   
The experimental bounds on the dimensionful coupling $\xi$ imply that we
must have ${\mbox{Re}}(\xi)
{\ \lower-1.2pt\vbox{\hbox{\rlap{$<$}\lower5pt\vbox{\hbox{$\sim$}}}}\ }
10^{-8}-10^{-7}~{\mbox{TeV}}^{-2}$ 
\cite{carone}\footnote{A rough estimate for the $K^0_L-K^0_S$ 
mass splitting due to (\ref{kaon}) is given by $\Delta m_K/m_K\sim
{\mbox{Re}}(\xi) m_K^2$, where $m_K$ is the Kaon mass. Note that
${\mbox{Im}}(\xi)$ is constrained by the CP violation parameters, and this
constraint is about 100 times stronger than that for ${\mbox{Re}}(\xi)$. We
will discuss this issue in section IV.}. Without
any flavor symmetries, we generically expect the above operator to be
suppressed at most by $1/M_s^2$. This would imply the following lower bound:
$M_s{\ \lower-1.2pt\vbox{\hbox{\rlap{$>$}\lower5pt\vbox{\hbox{$\sim$}}}}\ }
10^{3}-10^4~{\mbox{TeV}}$. To lower this bound on $M_s$ we would need to
impose some symmetry which acts on $s$ and $d$ states differently, that is,
we would have to impose a {\em flavor} symmetry \cite{flavor}. On the other
hand, to be compatible with, say, the observed fermion hierarchy in the
Standard Model, this flavor symmetry would have to be broken. So
suppressing FCNCs is a non-trivial task: the suppression should arise in a
subtle way due to a {\em broken} flavor symmetry. 

{}Next, we can ask what kind of flavor symmetries should be imposed to
possibly suppress the above operator. First, we should consider (continuous
or discrete) {\em gauge} flavor symmetries (see Introduction). Second, this
gauge symmetry must be anomaly free. Before we plunge into more technical
issues such as anomaly freedom, we can ask a more basic question: Would
Abelian flavor symmetries suffice, or should we impose non-Abelian flavor
symmetries? As pointed out in \cite{flavor}, Abelian symmetries alone
cannot do the job. Here we would like to briefly review the arguments 
of \cite{flavor}.    

\subsection{Possible Flavor Symmetries}

{}To understand why Abelian flavor symmetries are inadequate for
suppressing FCNCs in the present context, let us consider dimension 6
four-fermion operators containing the fields $Q_i$ and ${\overline 
Q}^{i}$, where the latter fields are just the conjugates of the former, and we
are using superscript for the corresponding flavor indices as ${\overline 
Q}^{i}$ transform in the representation of the flavor group which is
complex conjugate of the representation in which $Q_i$ transform. Here the
indices $i,j,\dots$ refer to
flavor indices in the {\em flavor} basis, that is, the basis in which the
charged electroweak currents are diagonal. Ultimately we will be interested
in understanding FCNCs in the {\em physical} basis in which the up and down
quark mass matrices are diagonal. These two bases are related by the usual
unitary rotations (see below). 

{}Thus, let us consider the most general four fermion operators of the
form: 
\begin{equation}
 C^{ij}_{{k}{l}} ({\overline Q}^{ i} Q_k)
 ({\overline Q}^{ j} Q_l)~.
\end{equation}
Note that no Abelian (continuous or discrete) symmetry 
can forbid the terms with
$i=k$ and $j=l$. Moreover, the corresponding coefficients
{\em a priori} are completely unconstrained. These terms do not contain the
dangerous operator $({\overline s}d)^2$ in the flavor basis. However, in
the physical basis this term will be generated unless the flavor and
physical quark states are identical, at least for the first two
generations. 
This could {\em a priori} be the case
in, say, the down quark sector. However, this {\em cannot} be the case in
{\em both} up and down quark sectors: indeed, if both up and down quark mass
matrices were diagonal in both bases
(or, more precisely, if this was the case for the
corresponding $1-2$ blocks, $1,2,3$ referring to the three generations),
there would be no Cabibbo mixing between the first and the second
generations, which would contradict the experimental data. This implies
that either $({\overline s}d)^2$ or $({\overline u}c)^2$ (or both at the
same time) operators would be present in the physical basis with the
coupling $\xi\sim \sin^2(\theta_C)/M_s^2$ ($\theta_C$ is the Cabibbo angle)
if the flavor symmetry is Abelian. This is unacceptable as the
corresponding strengths of flavor violation in the ${\overline K}^0-K^0$ and/or
${\overline D}^0-D^0$ transitions would be above the present experimental
bounds (note that $\sin \theta_C\simeq 0.22$). The only way around this
problem is to impose a non-Abelian flavor symmetry which is more
constraining and can {\em a priori} result in a non-trivial ``conspiracy''
between the coefficients $C^{ij}_{kl}$ such that the
disastrous flavor violating operators are not induced in the physical basis
\cite{flavor}.

{}We are therefore led to the conclusion that some type of non-Abelian
flavor symmetry must be invoked\footnote{Some early works on
non-Abelian flavor symmetries include \cite{fl-old}. Some of the more
recent works can be found in \cite{fl-new}.}. 
If all the Yukawa couplings in the
Standard Model are set to zero, then we have the following flavor symmetry
group: $G_F=\bigotimes_A U(3)_A$, where $A=Q,D,U,L,E,N$. This is the
largest flavor group we can attempt to introduce. However, it is clear that
at least in the quark sector the $U(3)_Q$ and $U(3)_U$ must be broken at
the string scale to have large top mass. In fact, this is also the case for
$U(3)_D$ as the top-bottom splitting is generally considered to be due to a
``vertical'' hierarchy: we can either have large $\tan \beta$, or the
$QDH_-$ coupling could arise as an effective Yukawa coupling descending
from a higher dimensional operator $SQDH_-$ upon the
additional singlet field $S$ acquiring a VEV $\langle S\rangle/M_s\sim
m_b/m_t$ (here $m_b$ and $m_t$ are the bottom and top quark masses,
respectively, and $\tan\beta$ is assumed to be $\sim 1$). 
As was pointed out in \cite{zura1}, the latter possibility
can naturally arise in the TSSM. The same is also expected to be the case
in the lepton sector. Thus, we should really consider the $U(2)_A$ subgroup
of the $U(3)_A$ flavor group as our starting point.

{}Thus, let us consider gauging as large a non-Abelian
flavor group as possible. In the next section we will argue that the anomaly 
cancellation conditions in the TSSM require that we identify $SU(2)_Q$ with
$SU(2)_L$ - we will refer to this flavor subgroup as the left-handed flavor
group $SU(2)_L$. Similarly, we must identify the other four subgroups
$SU(2)_D$, $SU(2)_U$, $SU(2)_E$ and $SU(2)_N$ with each other - we will 
refer to this
flavor subgroup as the right-handed flavor group $SU(2)_R$. We can also
gauge additional flavor $U(1)$'s (albeit the number of relevant anomaly free
possibilities is rather limited). We will consider such an Abelian flavor
group (as well as its discrete subgroups) in the next section in more
detail. Here, however, we would like to concentrate on the non-Abelian part
of the flavor group, the largest possibility being $SU(2)_L\otimes
SU(2)_R$.

{}Let us see if the $SU(2)_L\otimes SU(2)_R$ flavor symmetry can suppress
FCNCs adequately. Let us concentrate on the first and the second
generations for the moment. In particular, in the expressions we are about
to write down the flavor indices $i,j,\dots$   
take values $1,2$. Thus, consider four-fermion operators in the presence of
the $SU(2)_L\otimes SU(2)_R$ flavor symmetry. The $SU(2)_L\otimes SU(2)_R$
invariant operators involving only
the fields $Q_i$ and ${\overline Q}^{i}$ are given by 
\begin{equation}\label{Q}
 C ({\overline Q}^{i} Q_i)
 ({\overline Q}^{j} Q_j)+
 C^\prime\epsilon_{ij}\epsilon^{{k}{ l}} 
 ({\overline Q}^{i} Q_k)
 ({\overline Q}^{j} Q_l)~. 
\end{equation}
These operators do not contain the dangerous four-fermion interactions in
the flavor basis. However, we really need to go to the physical basis. This
is accomplished by means of the rotation $Q_i=
{{\cal Q}_i}^I Q_I$, which is unitary if we consider all three
generations. 
Here $Q_I$ contains the left-handed up and down quarks 
$(Q_U)_I$ respectively $(Q_D)_I$ in the physical basis, 
and by ${{\cal Q}_i}^I$ we really mean two {\em a
priori} independent matrices ${({\cal Q}_U)_i}^I$ and  
$({{\cal Q}_D)_i}^I$ acting on $(Q_U)_I$ respectively $(Q_D)_I$. The condition
that ensures that the dangerous operator $({\overline s}d)^2$ 
does not appear in the physical basis is then given by
the requirement that the $2\times 2$ matrix $({{\cal Q}_D)_i}^I$ 
be almost unitary.
A similar constraint arises if we consider the four-fermion operators
involving $Q$'s and/or $D$'s (as well as their conjugates): the $2\times 2$
matrix ${{\cal D}^{\bar I}}_{\bar i}$ must be almost
unitary. Here we are using the bar
notation to distinguish the $SU(2)_R$ flavor indices ${\bar i},{\bar I},\dots$ 
from the $SU(2)_L$ flavor indices $i,I,\dots$, and the left-handed down
anti-quarks in the physical basis are given by $D_{\bar I}$, where 
$D_{\bar i}=D_{\bar I}{{\cal D}^{\bar I}}_{\bar i}$. 
Finally, the $2\times 2$ matrix ${{\cal U}^{\bar I}}_{\bar i}$, which
relates the left-handed up anti-quarks in the flavor and physical bases via
$U_{\bar i}=U_{\bar I} {{\cal U}^{\bar I}}_{\bar i}$, must also be almost
unitary. (Note that 
$({{\cal Q}_D)_i}^I$ and ${{\cal D}^{\bar I}}_{\bar i}$ together 
diagonalize the down quark mass matrix $({\cal M}_D)^{{\bar i}i}$ via
$D_{\bar i}({\cal M}_D)^{{\bar i}i} (Q_D)_i=
D_{\bar I}({\cal M}_D)^{{\bar I}I} (Q_D)_I$, where the diagonal mass matrix
$({\cal M}_D)^{{\bar I}I}\equiv {{\cal D}^{\bar I}}_{\bar i}({\cal
M}_D)^{{\bar i} i}({{\cal Q}_D)_i}^I$. Similarly, 
$({{\cal Q}_U)_i}^I$ and ${{\cal U}^{\bar I}}_{\bar i}$ together 
diagonalize the up quark mass matrix $({\cal M}_U)^{{\bar i}i}$ via
$U_{\bar i}({\cal M}_U)^{{\bar i}i} (Q_U)_i=
U_{\bar I}({\cal M}_U)^{{\bar I}I} (Q_U)_I$, where the diagonal mass matrix
$({\cal M}_U)^{{\bar I}I}\equiv {{\cal U}^{\bar I}}_{\bar i}({\cal
M}_U)^{{\bar i} i}({{\cal Q}_U)_i}^I$. Here we are considering all three
generations so that all matrices are $3\times 3$ matrices.) 

{}The above ``unitarity'' constraints imply that the $1-3$ and $2-3$ mixing
in the down as well as the up quark sectors cannot be large. This
requirement, however, is not so constraining. Thus, if we take the 
$1-3$ and $2-3$ mixings in the up and down quark sectors to be of order of
the corresponding mixings in the Cabibbo-Kobayashi-Maskawa (CKM)
matrix $V_{\small{CKM}}\equiv {\cal
Q}_U^\dagger {\cal Q}_D$, then the FCNCs induced by the ``non-unitarity''
of the $1-2$ diagonalizations in the up and down quark sectors will be
adequately suppressed for $M_s\sim 10-100~{\mbox{TeV}}$. In particular, the
$1-3$ and $2-3$ CKM mixings are given by $|V_{ub}|\sim 3\times 10^{-3}$ and
$|V_{cb}|\sim 3\times 10^{-2}$. This implies that the corresponding flavor
violating couplings are going to be given by:
\begin{eqnarray}
 &&\xi_{({\overline s} d)^2}\sim \xi_{({\overline u} c)^2}
 \sim |V_{ub}V_{cb}|^2/M_s^2 \sim
 10^{-8}/M_s^2~,\nonumber\\ 
 &&\xi_{({\overline b} d)^2}\sim |V_{ub}|^2/M_s^2 \sim
 10^{-5}/M_s^2~,\nonumber\\
 &&\xi_{({\overline b} s)^2}\sim |V_{cb}|^2/M_s^2 \sim
 10^{-3}/M_s^2~.\nonumber
\end{eqnarray}
These couplings are within the experimental bounds for 
$M_s\sim 10-100~{\mbox{TeV}}$. 
   
\subsection{Gauge and Flavon Mediated Flavor Violation}

{}Imposing a flavor symmetry is not the end of the story, however, as it
has to be broken. As pointed out in \cite{flavor}, it cannot be broken on
the $p$-branes as this would be in contradiction with various cosmological
as well as other constraints \cite{flavon}. This implies that it has to be
broken in the bulk, and if we gauge a continuous flavor symmetry, it must
be a ``bulk'' gauge symmetry. Note that the flavor symmetry breaking source
could be on a ``distant'' brane \cite{flavor1}, but it would have to be
transmitted to the $p$-branes via some ``bulk'' fields anyways, 
so for our purposes
here we can treat the flavor symmetry breaking as arising due to the
corresponding ``bulk'' flavons acquiring non-zero VEVs. 

{}The flavor symmetry breaking due to the ``bulk'' flavons is felt by the
$p$-brane fields via the couplings of the latter to the former. The
non-trivial flavor hierarchy in the Standard Model then would have to be
due to the ``bulk'' flavons. It is then clear that once the ``bulk''
flavons acquire non-zero VEVs, there are going to be induced effective
flavor violating four-fermion operators. We will discuss the flavon
VEV-induced flavor violation in detail in section IV. Here, however, we we
will focus on the other two sources of flavor violation which we must
take into account. 

{}First, we must take into account the flavon mediated flavor
violation due to the flavon exchanges. Let $n$ be the number of large
transverse dimensions in which the flavons can propagate. Then it is
straightforward to estimate the corresponding flavor non-universal
contributions due to the
light (${\
\lower-1.2pt\vbox{\hbox{\rlap{$<$}\lower5pt\vbox{\hbox{$\sim$}}}}\ } 
m_M$) and heavy 
$({\ \lower-1.2pt\vbox{\hbox{\rlap{$>$}\lower5pt\vbox{\hbox{$\sim$}}}}\ }
m_M$) KK flavon modes in these directions ($m_M$ is the corresponding meson
mass, $M=K,D,B$) \cite{flavor}:
\begin{eqnarray}
 &&\xi_{light} \sim (\lambda M_s)^{-2} (m_M/M_s)^n~,\nonumber\\
 &&\xi_{heavy} \sim \lambda^{n-4} (m_M/M_s)^2~.\nonumber
\end{eqnarray}     
Here $\lambda\sim m_c/m_t\sim 10^{-2}$ ($m_c$ is the charm quark mass)
is a measure of the flavor symmetry breaking.
Thus, in the worst case where $n=2$ we have flavor violation which 
is in the acceptable range for  $M_s {\
\lower-1.2pt\vbox{\hbox{\rlap{$>$}\lower5pt\vbox{\hbox{$\sim$}}}}\ }
30~{\mbox{TeV}}$ or so.

{}Suppose, however, we have a continuous non-Abelian 
flavor gauge symmetry in the
bulk. Then we must also consider gauge mediated flavor violation due to the
exchanges of horizontal gauge bosons and their KK counterparts all of which
have flavor non-universal contributions ($\sim (\lambda M_s)^2$) to their 
squared masses. The induced flavor
violation is given by \cite{flavor}:  
\begin{eqnarray}
 &&\xi_{gauge} \sim \lambda^{n-2} M_s^{-2}~,~~~n<4~,\nonumber\\
 &&\xi_{gauge} \sim \lambda^2 M_s^{-2}~,~~~n\geq 4~.\nonumber
\end{eqnarray}
Note that for $n<4$
the gauge mediated flavor violation is unacceptably large. In fact, for
$n=2$ there is no suppression at all (which is 
due to the fact that in this case the flavor
non-universal contribution to the gauge boson mass squared scales the same
way with $\lambda$ as the number of the ``light'' KK modes for which this
contribution is dominant\footnote{By such ``light'' KK modes we mean those
for which the flavor {\em universal} mass squared contributions due 
to the non-zero KK momenta are ${\
\lower-1.2pt\vbox{\hbox{\rlap{$<$}\lower5pt\vbox{\hbox{$\sim$}}}}\ } 
(\lambda M_s)^2$.}). We are therefore led to the conclusion that the
FCNC constraints rule out gauging continuous flavor symmetries in the bulk
for $n<4$. In the next section we will circumvent this difficulty by
explicitly constructing a {\em discrete} flavor gauge 
symmetry that suppresses the
FCNCs adequately, yet has the advantage of not reintroducing gauge mediated
flavor violation as there are no gauge bosons in this case to begin with. 

\section{Discrete Flavor Gauge Symmetries and FCNC Suppression}

{}In this section we will explicitly construct an anomaly free non-Abelian
discrete 
flavor gauge symmetry which suppresses FCNCs just as well as the
$SU(2)_L\otimes SU(2)_R$ continuous flavor gauge symmetry. In fact, the
discrete flavor group we are going to discuss is a subgroup of
$SU(2)_L\otimes SU(2)_R$. Since we are going to gauge it, we must make sure
that all anomalies cancel. The discrete anomaly cancellation conditions are
similar to those for continuous gauge symmetries \cite{IR1}. To illustrate
some of the non-trivial issues arising when discussing discrete anomaly
cancellation conditions, let us first consider these conditions in the case
of an Abelian ${\bf Z}_N$ discrete gauge symmetry.  

{}A ${\bf Z}_N$ discrete gauge symmetry can be thought of as follows. 
Consider a theory with some matter charged under an anomaly free $U(1)$ 
gauge symmetry. Let all the $U(1)$ charge assignments be integer. 
Consider now adding a pair of chiral superfields which are neutral under 
all the other gauge subgroups but carry $+N$ and $-N$ charges under the 
above $U(1)$. Suppose there is a flat direction along which these chiral 
superfields acquire non-zero VEVs. Then the $U(1)$ gauge symmetry is broken 
down to its ${\bf Z}_N$ subgroup. This ${\bf Z}_N$ is then an anomaly free
discrete gauge symmetry. In the above approach the anomalies for the 
${\bf Z}_N$ gauge symmetry mimic those for the original $U(1)$ gauge 
symmetry (except that the former anomalies are only defined ``modulo
$N$''). Thus, we have the ${\mbox{Tr}}({\bf Z}_N^3)$ anomaly as well as 
the mixed ${\mbox{Tr}}({\bf Z}_N)$ gravitational anomaly. If there are 
{\em non-Abelian} gauge subgroups $\bigotimes_i G_i$ in the theory, one
also needs to consider the mixed ${\mbox{Tr}}(G_i^2 {\bf Z}_N)$ non-Abelian 
gauge anomalies. Finally, if there are additional {\em Abelian} subgroups 
$\bigotimes_a U(1)_a$, then we must also consider the mixed 
${\mbox{Tr}}(U(1)_a U(1)_b {\bf Z}_N)$ and ${\mbox{Tr}}(U(1)_a {\bf
Z}^2_N)$ gauge anomalies. The last two anomalies are somewhat tricky 
compared with the rest of the anomalies. The reason is that to compute 
them one is required to know the massive spectrum of the theory. More 
concretely, these anomalies depend on details of the parent $U(1)$ 
breaking \cite{IR1}. Since we are going to attempt to gauge discrete 
symmetries in the TSSM which contains an Abelian gauge subgroup 
(namely, $U(1)_Y$), we cannot ignore these anomalies. There is however, 
a way out of this difficulty. We can explicitly gauge a {\em continuous} 
(that is, $U(1)$) symmetry and make sure that all the anomalies 
cancel before we break it to the corresponding discrete subgroup. 
This way we are guaranteed to have an anomaly free discrete gauge symmetry 
at the end of the day.

{}Similar considerations apply to any discrete flavor gauge symmetry
$\Gamma_F$. To make sure that it is anomaly free, we can first consistently
gauge a continuous flavor symmetry $G_F$ which contains $\Gamma_F$ as a
subgroup (note that $G_F$ as well as $\Gamma_F$ can contain Abelian as well
as non-Abelian factors). Then the discrete flavor symmetry $\Gamma_F$ is
guaranteed to be anomaly free. This is precisely the strategy we will follow
in this section.   

\subsection{The $SU(2)_L\otimes SU(2)_R$ Flavor Gauge Symmetry}

{}As our starting point, we would like to consistently gauge the
$SU(2)_L\otimes SU(2)_R$ flavor symmetry. 
{\em A priori} we could start from the full flavor
group $G_F=\bigotimes_A U(3)_A$, where $A=Q,D,U,L,E,N$, where the 3
generations transform in the corresponding fundamental representations of
the subgroups $U(3)_A$. 
However, as we already mentioned in the previous section, we
need not bother trying to gauge 
the full $U(3)_A$ subgroups as they would have to be
broken at the string scale $M_s$ to account for the large quark and lepton
masses in the third generation. We can therefore restrict our attention to
the flavor group $G_F=\bigotimes_A U(2)_A$. Here we must address the issue
of the mixed ${\mbox{Tr}}(G_F^2\otimes U(1)_Y)$ gauge 
anomalies. These could {\em a
priori} be canceled by introducing additional states charged under
$G_F\otimes U(1)_Y$ (but neutral under $SU(3)_c\otimes SU(2)_w$). The
generic problem with this is that we would have to add many such states,
and they would have to be chiral (to be able to cancel anomalies). They
then are generically massless unless $U(1)_Y$ is broken, which only happens
upon the electroweak symmetry 
breaking\footnote{In fact, some of these additional states would actually
have to be $SU(2)_w$ doublets to have appropriate couplings with the
electroweak Higgs doublets $H_\pm$, which makes the anomaly cancellation
(as well as other issues) even more problematic.}. 
This then implies that there would
be many additional light states on top of those we already have in the
TSSM. Note that the situation is even worse once we consider the mixed 
${\mbox{Tr}}(G_F^2\otimes ({\widetilde {\cal L}}_3\otimes {\widetilde {\cal
R}}_3))$ gauge anomalies, where ${\widetilde {\cal L}}_3\otimes
{\widetilde {\cal R}}_3$ is the Abelian discrete gauge symmetry discussed
in section II which is responsible for proton stabilization.

{}The above difficulties can be ameliorated by considering the flavor
group 
$G_F=SU(2)_L\otimes SU(2)_R$, where we identify $SU(2)_A$, $A=Q,L$, with the
left-handed flavor group $SU(2)_L$, and $SU(2)_A$, $A=D,U,E,N$, with the
right-handed flavor group $SU(2)_R$. {\em A priori} we could also consider
additional $U(1)$ factors, but the anomaly cancellation conditions are very
tight, so the number of consistent possibilities is actually very
limited. We will consider adding such a $U(1)$ factor in a moment. First,
however, let us discuss gauging $G_F=SU(2)_L\otimes SU(2)_R$.

{}Next, we give the quantum numbers of the TSSM fields under 
$[SU(2)_L\otimes SU(2)_R]\otimes [U(1)_{\cal L}\otimes U(1)_{\cal R}]$ (here
$U(1)_{\cal L}$ and $U(1)_{\cal R}$ are the ``parent'' $U(1)$
symmetries\footnote{Here we choose to work with these parent gauge
symmetries as the mixed anomalies are more transparent in this language.}
for the discrete subgroups ${\widetilde {\cal L}}_3$ respectively
${\widetilde {\cal R}}_3$):   
\begin{eqnarray}
 &&Q_i:~({\bf 2},{\bf 1})(0,0)~,~~~Q_3:~({\bf 1},{\bf 1})(0,0)~,\nonumber\\
 &&D_{\bar i}:~({\bf 1},{\bf 2})(0,+1)~,~~~
 D_{\bar 3}:~({\bf 1},{\bf 1})(0,+1)~,\nonumber\\
 &&U_{\bar i}:~({\bf 1},{\bf 2})(0,-1)~,~~~
 U_{\bar 3}:~({\bf 1},{\bf 1})(0,-1)~,\nonumber\\
 &&L_i:~({\bf 2},{\bf 1})(+1,0)~,~~~L_3:~({\bf 1},{\bf 1})(+1,0)~,\nonumber\\
 &&E_{\bar i}:~({\bf 1},{\bf 2})(-1,+1)~,~~~
 E_{\bar 3}:~({\bf 1},{\bf 1})(-1,+1)~,\nonumber\\
 &&N_{\bar i}:~({\bf 1},{\bf 2})(-1,-1)~,~~~
 N_{\bar 3}:~({\bf 1},{\bf 1})(-1,-1)~,\nonumber\\
 &&\chi_{i\alpha}:~({\bf 2},{\bf 1})(-1,0)~,~~~
 \chi^\prime_{{\bar i}\alpha^\prime}:~({\bf 1},{\bf 2})(+1,0)~.\nonumber
\end{eqnarray} 
Here the $SU(2)_L\otimes SU(2)_R$ quantum numbers are given in bold font,
whereas the $U(1)_{\cal L}\otimes U(1)_{\cal R}$ charges are given in
parentheses. The TSSM states $H_\pm$ and $F_\pm$ are neutral under 
$SU(2)_L\otimes SU(2)_R$, and are not shown. The flavor indices $i$ and ${\bar
i}$ take values $1,2$ and ${\bar 1},{\bar 2}$, respectively.
The new states $\chi_{\alpha}$ and $\chi^\prime_{\alpha^\prime}$ 
($\alpha,\alpha^\prime=1,2$ are new indices {\em not} related to the flavor
indices for the quarks and leptons) are
neutral under the Standard Model gauge group $SU(3)_c\otimes SU(2)_w\otimes
U(1)_Y$, and they are required by the ${\mbox{Tr}}(SU(2)^2_L U(1)_{\cal
L})$ and ${\mbox{Tr}}(SU(2)^2_R U(1)_{\cal L})$ mixed gauge 
anomaly cancellation,
respectively. Note that these new states carry non-zero $U(1)_{\cal L}$
charges. This could be dangerous for proton stability which is based on a
non-trivial selection rule (due to the ${\widetilde {\cal L}}_3\otimes 
{\widetilde {\cal R}}_3$ discrete gauge symmetry) 
discussed in detail in \cite{zura1}. To avoid jeopardizing proton
stability, we can simply require that the new states
$\chi_\alpha,\chi^\prime_{\alpha^\prime}$ be heavier than
proton (and their VEVs are zero). This can be achieved by introducing 
$n_\Sigma$ 
additional ``flavon'' fields (which are neutral under $SU(3)_c\otimes
SU(2)_w\otimes U(1)_Y$) 
\begin{equation} 
 \Sigma^{i{\bar i}}_a\:~({\bf 2},{\bf 2})(0,0)~,~~~a=1,\dots,n_\Sigma~,
\end{equation}
with the following Yukawa couplings to 
$\chi_{\alpha},\chi^\prime_{\alpha^\prime}$:
\begin{equation} 
 y_{a\alpha\alpha^\prime} 
 \Sigma^{i{\bar i}}_a \chi_{i\alpha}\chi^\prime_{{\bar i}\alpha^\prime}~.
\end{equation}
Upon the fields $\Sigma_a$ (which we assume to be ``bulk'' flavon fields) 
acquiring
VEVs (which break the $SU(2)_L\otimes SU(2)_R$ flavor symmetry), 
the fields
$\chi_\alpha,\chi^\prime_{\alpha^\prime}$ pick up large masses (of order of
the flavor symmetry breaking scale which is $\sim \lambda M_s\sim
100~{\mbox{GeV}}-1~{\mbox{TeV}}$, where $\lambda\sim 10^{-2}$ was
introduced in the previous section), so that proton decay via channels
involving $\chi_\alpha$ and/or $\chi^\prime_{\alpha^\prime}$ in the final state
is forbidden due to the kinematics. Note that such heavy fields could {\em
a priori} be dangerous - they could potentially 
mediate unacceptably large FCNCs
through a tree-level exchange. However, precisely due to their non-zero 
$U(1)_{\cal L}$ (or, more precisely, ${\widetilde{\cal L}}_3$) charges, 
the $\chi_\alpha,\chi^\prime_{\alpha^\prime}$ states
do not have the required couplings (such as, say, $\chi QDH_-$) to the
quarks (or leptons) to mediate flavor violation in this way, 
and are therefore safe.   

{}The above spectrum (together with the states $H_\pm,F_\pm$) is completely
anomaly free. This implies that if we considered the same
spectrum\footnote{At the end of the day we are going to consider the
discrete subgroup ${\widetilde {\cal L}}_3\otimes{\widetilde {\cal R}}_3$
of the $U(1)_{\cal L}\otimes U(1)_{\cal R}$ gauge group as in \cite{zura1}.}
with the discrete flavor gauge symmetry $\Gamma_L\otimes \Gamma_R$ 
(instead of the continuous flavor gauge symmetry $SU(2)_L\otimes SU(2)_R$),
where $\Gamma\subset SU(2)$, the resulting spectrum would also be completely
anomaly free. In the next subsection we will make a particular choice of
$\Gamma$ such that in the unbroken flavor symmetry limit the FCNCs are
suppressed just as well as in the case of the full $SU(2)_L\otimes SU(2)_R$ 
flavor gauge symmetry.

\subsection{A Conclusive Solution: The $T_L\otimes T_R$ Discrete Flavor
Gauge Symmetry}

{}Next, we need to identify a discrete subgroup $\Gamma$ of $SU(2)$ such
that the $\Gamma_L\otimes \Gamma_R$ discrete flavor gauge symmetry
suppresses FCNCs just as well as the $SU(2)_L\otimes SU(2)_R$ flavor gauge
symmetry itself. It is not difficult to identify such a subgroup - 
the discrete subgroups of $SU(2)$ are classified in terms of the A-D-E
series. The infinite $A_N$ series corresponds to the Abelian ${\bf Z}_N$
subgroups of $SU(2)$. The second infinite $D_N$ series contains non-Abelian
dihedral groups (with $D_3$ isomorphic to (the double cover of) the 
permutation group $S_3$). Finally, the finite E-series contains three
``exceptional'' non-Abelian subgroups (analogous to $E_6,E_7,E_8$): the
tetrahedral group $T$, the octahedral group $O$, and the icosahedral group
$I$. Here we are interested in the non-Abelian groups only. It is not
difficult to show that the dihedral
groups $D_N$ cannot do the job. The tetrahedral
group $T$, however, is perfectly adequate for our purposes. Thus,
it only allows the $SU(2)$ invariant four-fermion
operators. For instance, the only operators containing $Q$'s and
${\overline Q}$'s are those given in (\ref{Q}). This follows from the 
action of the generators of the
tetrahedral group $T$ on the fields $Q_i,Q_3$ ($i=1,2$). Let these
generators be $\theta,R,R^\prime$. Then their
action on $Q_3$ is trivial. The fields $Q_i$, however, transform in one,
namely, ${\bf 2}_0$, of
the three non-trivial two dimensional representations ${\bf 2}_k$,
$k=0,1,2$, of $T$. These two
dimensional representations are given by 
\begin{eqnarray}     
 &&\theta={\omega^{k}\over 2}({\bf
 1}-i\sigma_1+i\sigma_2-i\sigma_3)~,\nonumber\\ 
 &&R=i\sigma_1,~~~ R^\prime=i\sigma_3~,\nonumber 
\end{eqnarray}
where ${\bf 1}$ stands for the $2\times 2$ identity matrix, and
$\sigma_1,\sigma_2,\sigma_3$ are the Pauli matrices. Also,
$\omega\equiv \exp(2\pi i/3)$.
Note that
$\theta^3=R^2=R^{\prime 2}=-{\bf 1}$, and $R\theta=\theta R^\prime R$,
$R^\prime \theta =\theta R$, $R^\prime R\theta =\theta R^\prime$, and
$R^\prime R R^\prime =R$. (These are the defining commutation relations 
for the tetrahedral group.) Note that we
have just described the action of $T_L$ (on $Q$'s and $L$'s). The action of
$T_R$ (on $D$'s, $U$'s, $E$'s and $N$'s) is similar. 

{}Thus, the $T_L\otimes T_R$ discrete flavor gauge symmetry is just as
efficient in constraining four-fermion operators as the continuous
$SU(2)_L\otimes SU(2)_R$ flavor gauge symmetry whose subgroup $T_L\otimes
T_R$ is. This discrete symmetry, however, does not guarantee that {\em all}
dangerous flavor violating operators are suppressed {\em after} the flavor
symmetry breaking (neither does $SU(2)_L\otimes
SU(2)_R$). In the following  we will consider an additional Abelian
flavor symmetry which is required to ensure that the flavon VEV-induced
flavor violation is also adequately suppressed.  
     
\subsection{Suppressing Flavon VEV-induced FCNCs: The $U(1)_V$
Flavor Gauge Symmetry}

{}The $SU(2)_L\otimes SU(2)_R$ flavor gauge symmetry (or its discrete
subgroup $T_L\otimes T_R$) suppresses dangerous flavor violating
four-fermion operators {\em before} the flavor symmetry breaking takes
place. Once the flavor symmetry is broken, there are going to be induced 
such four-fermion operators, but the corresponding couplings will be
suppressed by the flavon VEVs. Nonetheless, we must make sure that this
suppression is adequate, that is, that these couplings are within the
experimental bounds. In this respect it does not make any difference
whether we are breaking the $SU(2)_L\otimes SU(2)_R$ flavor gauge symmetry
or its discrete subgroup $T_L\otimes T_R$, so we will use the language of
the full $SU(2)_L\otimes SU(2)_R$ flavor gauge symmetry in this subsection.    

{}Let us first consider the operators involving only the first two
generation quarks and anti-quarks in the {\em flavor} basis. The possible
VEV-induced operators of this type are always suppressed by two powers of
$\lambda\sim m_c/m_t$. To see this, note that, say, in the operator
$(D_{\bar i} Q_i)({\overline Q}^j {\overline D}^{\bar j})$ 
we must contract two pairs of
the $SU(2)$ indices\footnote{Note that operators of, say,  the form 
$(D_{\bar i} Q_i)(D_{\bar j}Q_j)$ are further suppressed by a factor
$\sim (\langle H_-\rangle/M_s)^2\sim (M_{ew}/M_s)^2$ 
as they carry two units of the weak isospin. 
In the following we will therefore not discuss such operators, but rather 
focus on ``self-conjugate'' operators of the type mentioned above.}. 
Let us first consider the flavon VEV-induced operators
of this type involving only flavons in the bifundamental representation of
$SU(2)_L\otimes SU(2)_R$, say, the fields $\Sigma^{i{\bar i}}_a$ introduced
above. The corresponding four-fermion operators then read:
\begin{equation}
 (D_{\bar i} Q_i)({\overline Q}^j {\overline D}^{\bar j} )
 \epsilon_{jk} \epsilon_{{\bar j}{\bar k}}\left(C_{a_1 a_2}
 \Sigma^{i{\bar i}}_{a_1}
 \Sigma^{k{\bar k}}_{a_2} + 
 C^\prime_{a_1 a_2}
 \Sigma^{i{\bar k}}_{a_1}
 \Sigma^{k{\bar i}}_{a_2}\right)~,
\end{equation}  
where $C_{a_1 a_2}$ and $C^\prime_{a_1 a_2}$
are some dimensionful couplings of order $\sim 1/
M_s^4$. The strength of the induced FCNC operator of the type $({\overline
s}d)^2$ is then going to be suppressed at least as 
$\xi_{({\overline s}d)^2}\sim \lambda^2
/M_s^2$. Here we are taking into account that the $\Sigma^{i{\bar i}}_{a}$
VEVs are expected to be such that $\Sigma^{2{\bar 2}}_{a}\sim \lambda M_s$
to explain the flavor hierarchy between the charm mass $m_c$ and the top
mass $m_t$. The above coupling then follows if we assume the maximal mixing 
between the first and the second generations. However, in the case of
bifundamental flavons there is
going to be an additional suppression factor 
as the first and the second
generation mixing is expected to be somewhat smaller, and it is reasonable
to assume that it is (roughly) 
given by the Cabibbo angle $\theta_C$. Then the
corresponding coupling is going to be even more suppressed: 
$\xi_{({\overline s}d)^2}\sim \lambda^2 \sin^2(\theta_C)
/M_s^2$. In fact, to avoid fine-tuning in the determinants of the $1-2$
blocks in the up and/or 
down quark mass matrices, we can assume that $\Sigma^{1{\bar
2}}_a,\Sigma^{2{\bar 1}}_a\sim \lambda \sin(\theta_C) M_s$, which would lead
to the same conclusion for the coupling $\xi_{({\overline s}d)^2}$. (We
will discuss the up and down quark mass matrices in more detail in
the next subsection.) The above suppression is completely adequate for 
$M_s\sim 10-100~{\mbox{TeV}}$. A similar conclusion also holds for flavons
transforming in the fundamental representations. 

{}There are, however, other operators we must worry about which generically
are not as suppressed. These are the operators involving the third
generation quarks or anti-quarks. Thus, consider the following operator:  
\begin{equation}
 C_{a} ({\overline Q}^3 Q_i)(D_{\bar 3} {\overline D}^{\bar j})
 \Sigma^{i{\bar i}}_{a} \epsilon_{{\bar i}{\bar j}}~,
\end{equation}
where the dimensionful coupling $C_a\sim 1/M_s^3$. It is then not difficult
to see that the VEV-induced four-fermion operators of the type 
$({\overline b} d)^2$ (and $({\overline b} s)^2$) will be suppressed only by
the couplings of order $\xi\sim\lambda\sin(\theta_C)/M^2_s$. 
However, the above operator can be further suppressed 
by imposing an additional ``vector-like'' flavor symmetry that
acts non-trivially, say, on the third generation quarks and anti-quarks
only. In fact, even an Abelian symmetry would suffice for this purpose. We
can therefore augment the $SU(2)_L\otimes SU(2)_R$ flavor gauge symmetry by 
such a $U(1)_V$ flavor gauge symmetry. Thus, let the $U(1)_V$ charge
assignments be the following: $+1$ for $Q_3$, and $-1$ for $D_{\bar 3}$ and
$U_{\bar 3}$. Then it is clear that the above coupling is going to be
suppressed. (Note that the flavon $\Sigma^{i{\bar i}}_{a}$ must be neutral
under $U(1)_V$ to have the required couplings with the first and the second
generations.) In fact, all the other flavon VEV-induced FCNCs can be
suppressed this way. Thus, for instance, the operators of the type 
\begin{equation}
 ({\overline Q}^3 Q_i)({\overline Q}^3 Q_j)~,
\end{equation}
where the $SU(2)_L$ indices must be contracted with flavons transforming in
the fundamental representation, are also suppressed provided that we can
find an appropriate $U(1)_V$ charge assignment for the 
flavons\footnote{Here we should point out that the flavon VEV-induced FCNCs
due to the flavons transforming in the fundamental representations can be
suppressed even more if we utilize the vertical (that is, up-down)
hierarchy. Thus, consider a scenario where the $1-3$ and $2-3$ mixing
arises only in the down quark sector, and the flavons transforming in the
fundamental representations responsible for this mixing carry non-zero
(namely, $+3$) $U(1)_{\cal L}$ charges (the corresponding electroweak Higgs
doublet $H_-$ has $U(1)_{\cal L}$ charge $-3$ - see \cite{zura1} for
details). This implies that the corresponding flavon VEV-induced FCNCs are
suppressed by extra factors of $(m_b/m_t)^2$ (note that $\Lambda/M_s\sim 
m_b/m_t$, where $\Lambda$ is the $U(1)_{\cal L}\supset {\widetilde {\cal
L}}_3$ breaking scale \cite{zura1}).}. 

{}It is, however, non-trivial to construct an anomaly free
$U(1)_V$ 
flavor gauge symmetry. The anomaly cancellation constraints are
very tight as we have to consider mixed anomalies involving $SU(2)_L\otimes
SU(2)_R$, $SU(3)_c\otimes SU(2)_w\otimes U(1)_Y$, and $U(1)_{\cal L}\otimes
U(1)_{\cal R}$ (or, more precisely, the discrete ${\widetilde{\cal
L}}_3\otimes {\widetilde{\cal R}}_3 $ subgroup of the latter). Fortunately,
there exists a simple $U(1)_V$ flavor gauge symmetry 
which is anomaly free. Here we will first write down the $U(1)_V$ 
charge assignments for the
quarks and leptons, and then explain why this choice turns out to be
anomaly free. 

{}Thus, consider the following $U(1)_V$ 
charge assignments (the
$SU(2)_L\otimes SU(2)_R$ quantum numbers are given in bold font, whereas
the $U(1)_V$ 
charges are given in parenthesis):
\begin{eqnarray}
 &&Q_i:~({\bf 2},{\bf 1})(0)~,~~~Q_3:~({\bf 1},{\bf 1})(+1)~,
 \nonumber\\
 &&D_{\bar i}:~({\bf 1},{\bf 2})(0)~,~~~
 D_{\bar 3}:~({\bf 1},{\bf 1})(-1)~,\nonumber\\
 &&U_{\bar i}:~({\bf 1},{\bf 2})(0)~,~~~
 U_{\bar 3}:~({\bf 1},{\bf 1})(-1)~,\nonumber\\
 &&L_i:~({\bf 2},{\bf 1})(0)~,~~~L_3:~({\bf 1},{\bf 1})(-3)~,\nonumber\\
 &&E_{\bar i}:~({\bf 1},{\bf 2})(0)~,~~~
 E_{\bar 3}:~({\bf 1},{\bf 1})(+3)~,\nonumber\\
 &&N_{\bar i}:~({\bf 1},{\bf 2})(0)~,~~~
 N_{\bar 3}:~({\bf 1},{\bf 1})(+3)~,\nonumber\\ 
 &&\chi_{i\alpha}:~({\bf 2},{\bf 1})(0)~,~~~
 \chi^\prime_{{\bar i}\alpha^\prime}:~({\bf 1},{\bf 2})(0)~,\nonumber\\
 &&\Sigma^{i{\bar i}}_a:~({\bf 2},{\bf 2})(0)~,\nonumber\\
 &&\rho^i_\beta:~({\bf 2},{\bf 1})(+1)~,~~~
 (\rho^\prime_{\beta^\prime})^{\bar i}:~({\bf 1},{\bf 2})(-1)~,\nonumber\\
 &&{\eta}^i:~({\bf 2},{\bf 1})(-3)~,~~~
 ({\eta}^\prime)^{\bar i}:~({\bf 1},{\bf 2})(+3)~.\nonumber
\end{eqnarray}
All the other fields (such as $H_\pm,F_\pm$) are neutral under the $U(1)_V$
flavor gauge symmetry. On the other hand, the fields
($\beta,\beta^\prime=1,2,3$) 
$\rho_\beta, \rho^\prime_{\beta^\prime},\eta,\eta^\prime$ (as well as
$\Sigma_a$)  are all neutral under 
$SU(3)_c\otimes SU(2)_w\otimes
U(1)_Y\otimes {\widetilde {\cal L}}_3\otimes {\widetilde {\cal R}}_3$. The
above spectrum is completely anomaly free. This can be seen as
follows. Thus, the $U(1)_V$ symmetry acts as 3 times $B-L$ 
(baryon minus lepton number) for the third generation only (in the flavor
basis). This is precisely the reason why it is anomaly free and compatible
with all the other gauge symmetries we have introduced so far in the
TSSM. Thus, as was explained at length in \cite{zura1}, we can consistently 
gauge any linear combination of the three $U(1)$'s, namely,
$U(1)_Y,U(1)_{\cal L},U(1)_{\cal R}$. In fact, any linear combination of 
$U(1)_Y$ and $U(1)_{\cal R}$ can be gauged generation-by-generation. The
linear combination $V\equiv 3Y-3{\cal R}$ is precisely the generator
corresponding to 3 times $B-L$, which we choose to act on the third
generation only. In particular, the anomaly
cancellation requires that the
$U(1)_V$ charge assignments be different in the
quark and lepton sectors. Note that the fields 
$\rho_\beta, \rho^\prime_{\beta^\prime},\eta,\eta^\prime$
are chosen so that the corresponding $U(1)_V$ anomalies cancel. 

{}It is not difficult to check that with the above $U(1)_V$ charge
assignments all the flavon VEV-induced FCNCs are adequately suppressed. In
fact, the corresponding couplings are at least as suppressed as those due
to the ``non-unitarity'' of the $1-2$ diagonalization discussed in
subsection B of section III. Here we would like to point out that
suppressing flavon VEV-induced FCNCs
in the up and down quark sectors does not require the full $U(1)_V$
symmetry. More concretely, any discrete ${\bf Z}_N$ subgroup of $U(1)_V$ 
with $N\not=2,4$ would do the job. The $N=4$ case is also acceptable if the
$U(1)_V \supset {\bf Z}_4$ breaking scale is $10^{-2} M_s$ or 
lower\footnote{One of the
interesting phenomenological implications of the $U(1)_V$ breaking would be the
existence of new sub-millimeter forces which would compete with gravity
\cite{TeVphen}, and could be accessible at the upcoming sub-millimeter
experiments \cite{Price}. Thus, $U(1)_V$ must 
be a ``bulk'' gauge symmetry since
``bulk'' flavons are charged under it. This implies that upon breaking
$U(1)_V$ on a brane, 
the corresponding gauge $U(1)_V$ boson acquires the
mass around several inverse millimeters or so \cite{TeVphen}.}.  

{}Next, we would like to briefly discuss the flavor hierarchy in the above
model. 
Note that the couplings $Q_3 U_{\bar 3} H_+$ and $Q_3 D_{\bar 3} H_-$ are
allowed by the $G_F=SU(2)_L\otimes SU(2)_R \otimes U(1)_V$ 
symmetry. (This, in particular, is one of the reasons why we have chosen the
$U(1)_V$ symmetry to be ``vector-like''.) 
This implies
that we have large top and bottom quark masses with the ``vertical''
hierarchy generated as in \cite{zura1}. On the other hand, the first two
generations can only couple via 
\begin{eqnarray} 
 &&z_b H_+ Q_i U_{\bar i} \Sigma^{i{\bar i}}_a~,\nonumber\\
 &&{\widetilde z}_a 
 H_- Q_i D_{\bar i} {\Sigma}_a^{i{\bar i}}~,\nonumber
\end{eqnarray} 
where the dimensionful couplings $z_a\sim 1/M_s$ and 
${\widetilde z}_a\sim (m_b/m_t)z_a$ 
(here we are taking into account the ``vertical'' hierarchy mentioned
above). Upon the fields $\Sigma_a$  
acquiring non-zero
VEVs, we can generate non-zero masses for the second and first
generations\footnote{Here 
we are not going to consider the detailed mechanism of
how the VEVs for $\Sigma_a$ are generated to give a
desirable flavor hierarchy. This is a very model dependent question, which
needs to be addressed in any extension of the Standard Model. Let us,
however, mention that there are more than one possibilities for generating
such a hierarchy. Thus, for instance, 
the flavor symmetry could be broken via some dynamical mechanism on 
``distant'' branes \cite{flavor1}. Then the flavor hierarchy in the
``observable'' sector is determined by a model dependent flavor symmetry
breaking dynamics on these branes as well as their locations relative to
the $p$-branes on which the quarks and leptons are localized.}. In
particular, we can consider scenarios where the $1-2$ mixing occurs in both
the up and down quark sectors, or mostly in, say, the down quark sector.  
      
{}The $1-3$ and $2-3$ mixing arises due to the corresponding couplings to
the flavons $\rho^i_\beta$ and $(\rho^\prime_{\beta^\prime})^{\bar i}$:
\begin{eqnarray}
 &&v_\beta H_+ Q_i U_{\bar 3} \rho_\beta^i~,~~~w_\beta H_- Q_i D_{\bar 3} 
 \rho_\beta^i~,\nonumber\\
 &&v_{\beta^\prime}  H_+
 U_{\bar i} Q_3 (\rho^\prime_{\beta^\prime})^{\bar i}~,~~~
 w_{\beta^\prime}  H_- D_{\bar i} Q_3 (\rho^\prime_{\beta^\prime})^{\bar i}~,
 \nonumber
\end{eqnarray}
where $v_\beta, v_{\beta^\prime}\sim 1/M_s$, and $w_\beta, w_{\beta^\prime}\sim
(m_b/m_t)/M_s$.  
As we discussed in section III, we will assume that the
$\rho_\beta,\rho^\prime_{\beta^\prime}$ VEVs give rise to the $1-3$ and $2-3$
mixings comparable with those in the CKM matrix. This is compatible with
the observed flavor hierarchy and the CKM matrix (for a recent discussion,
see,{\em e.g.}, \cite{chka}, and references therein).  

{}Before we end this subsection, we would like to comment on CP
violation. Up until now we have been implicitly discussing the real parts of
the dimensionful couplings $\xi$ corresponding to flavor violating
four-fermion operators. The experimental bounds on the imaginary part of,
say, $\xi_{({\overline s} d)^2}$ are about 100 times stronger than those on
the real part. Here we would like to address the issue of whether these
imaginary flavor violating couplings are adequately suppressed as well. 

{}Let us first discuss the $\xi_{({\overline s} d)^2}$ and
$\xi_{({\overline u} c)^2}$ couplings. Note that in the absence of the
$1-3$ and $2-3$ mixing the up and down quark mass matrices are block-diagonal,
and the CP violation is absent. Thus, in this limit we expect the imaginary
parts of the corresponding flavor violating couplings to be vanishing. This
implies that the CP violating imaginary couplings are suppressed by the
$1-3$ and $2-3$ mixing angles. In fact, for these particular couplings this
suppression can be seen to be adequate for $M_s\sim 10-100~{\mbox{TeV}}$. 

{}Another coupling we must consider is  
${\mbox{Im}}(\xi_{({\overline b} d)^2})$. {\em A priori} this coupling is 
{\em not} suppressed any more than the corresponding real part. However,
suppose that the CP violating phases in the up and down quark mass matrices
are exactly equal (or very close to) $\pi/2$ (that is, we have the maximal
CP violation). When translated into the corresponding four-fermion
couplings, these phases are squared, so that the imaginary parts of these
couplings are vanishing. Thus, the maximal CP violation may provide a
framework for suppressing the corresponding flavor violating couplings.        

\subsection{The Lepton Sector}

{}So far we have discussed the quark sector of the model. The lepton sector
deserves a separate consideration as the $U(1)_V$ charge assignments here
are different from those in the quark sector. In particular,  
relatively large mixing angles between the first two and the third lepton
generations might be desirable in the light of \cite{Kami} (for a
recent discussion, see, {\em e.g}, \cite{Barb}, and references therein). 
In principle there appears to be no 
difficulty in obtaining such large mixing angles
as the corresponding flavons $\eta,\eta^\prime$ can acquire VEVs independent
from those in the quark sector.  

{}Next, we would like discuss the following
possibility. As was pointed out in \cite{zura1}, the ${\widetilde{\cal
L}}_3\otimes{\widetilde{\cal R}}_3$ discrete gauge symmetry is too strong for
just proton stabilization purposes. Thus, its subgroup ${\widetilde {\cal
Y}}_3$ is just as efficient. This subgroup can be viewed as the ${\bf Z}_3$
subgroup of the $U(1)_{\cal Y}$ gauge symmetry, where the generator ${\cal
Y}$ is given by ${\cal Y}={\cal L}-{\cal R}$. The advantage of having the
full ${\widetilde{\cal
L}}_3\otimes{\widetilde{\cal R}}_3$ discrete symmetry is that it
automatically forbids the dangerous dimension 5 operator $LLH_+H_+$ which
generically would result in too large Majorana neutrino masses (see section
II for details). {\em A priori} there is a possibility, however, that this
operator is suppressed due to the non-Abelian flavor gauge symmetries we
have been considering in this paper. This possibility was originally
pointed out in \cite{neutrino}, and also briefly discussed 
in \cite{zura1}. Now we
are in the position to see whether this operator is indeed sufficiently
suppressed in the context of a particular model we are considering in this
paper. 

{}It is convenient to discuss this issue using the language of the
$SU(2)_L\otimes SU(2)_R\otimes U(1)_V$ 
flavor gauge symmetry (as the conclusions are the
same for the $T_L\otimes T_R$ discrete flavor gauge symmetry). Let us first
consider the unbroken flavor symmetry limit.
Note that the $SU(2)_L\otimes SU(2)_R$ invariant operator
$\epsilon^{ij} L_i L_j H_+ H_+$ vanishes due to antisymmetry. On the other
hand, the $L_3 L_3 H_+ H_+$ operator is forbidden by the $U(1)_V$ flavor
symmetry. In fact, even if we confine our attention to its ${\bf Z}_N$
subgroup with $N\geq3$, this latter operator is still absent. 

{}Let us now consider the $SU(2)_L\otimes SU(2)_R$ breaking by the flavon
VEVs. Then {\em a priori} 
we are going to have VEV-induced operators of the form
$C^{ij}L_i L_j H_+ H_+$. Thus, for instance, we can have such operators 
with $C^{ij}=\eta^i{\overline \eta}_k \epsilon^{kj}$. (Here we could also
consider the $\rho_\beta$ flavons instead of the $\eta$ flavons.)
To suppress such
operators we can introduce a $U(1)_{V^\prime}$ flavor gauge symmetry (or
its appropriate discrete ${\bf Z}_N$ subgroup), where the vector-like
$U(1)_{V^\prime}$ symmetry acts as 3 times $B-L$ on the first two
generations only. However, then we have a ``generation-blind'' subgroup
$U(1)_{\cal Z}$ of
$U(1)_V\otimes U(1)_{V^\prime}$ that acts as 3 times $B-L$ on all three
generations. As was explained in detail in \cite{zura1}, gauging
$U(1)_{\cal Y}$ together with such a $U(1)_{\cal Z}$ gauge symmetry is
equivalent (for our purposes here) to gauging the full $U(1)_{\cal L}\otimes
U(1)_{\cal R}$ symmetry. 
In particular, note that ${\cal Z}=3Y-3{\cal R}$, where
${\cal Z}$ acts on all three generations. (Here the generalization to the
corresponding discrete subgroups should be evident.) In fact, it appears to
be the case that we need {\em two} generation-blind discrete gauge symmetries
to both ensure proton longevity and suppress the $LLH_+H_+$ operator.  

{}The above discussion might have interesting phenomenological
implications. Note that the ${\widetilde {\cal Y}}_3$ discrete gauge
symmetry allows all lepton number violating dimension 3 and 4 couplings in
the TSSM \cite{zura1}, whereas the full ${\widetilde{\cal
L}}_3\otimes{\widetilde{\cal R}}_3$ discrete gauge symmetry forbids all
such couplings. On the other hand, in the
MSSM with high $M_s$ (of order $M_{\small{GUT}}$) the only Abelian
generation-blind discrete gauge symmetry that forbids the dimension 5
operator $QQQL$ (which would otherwise be disastrous for proton stability
even in the case $M_s\sim M_{\small{GUT}}$) {\em and} at the same time
allows the $LLH_+H_+$ operator (which in this case is needed to generate
the correct neutrino masses via the old ``see-saw'' mechanism
\cite{see-saw}) is the ${\widetilde {\cal Y}}_3$ discrete gauge
symmetry \cite{IR0,zura1}. This would lead to an interesting experimental
``prediction'' that if $M_s$ is high (that is, $\sim M_{\small{GUT}}$ or so),
the upcoming collider experiments should detect lepton number violating
dimension 4 operators via channels involving sleptons. On the other hand,
if $M_s$ is low (that is, in the TeV range), such operators should be
absent. This implies that it might indeed be possible to
indirectly deduce whether $M_s$ is high or low by examining the
corresponding processes at the upcoming collider experiments {\em without}
directly producing the heavy Kaluza-Klein or string states. This might be
important as the nearest future collider experiments may not be able to
directly probe such states if they are heavier than a few TeV.

\section{Summary and Open Questions}

{}Let us briefly summarize the discussions in the previous sections. We have
considered the issue of FCNC suppression in the TSSM. We have shown that
the $T_L\otimes T_R$ non-Abelian discrete flavor gauge symmetry is just as
efficient in suppressing the corresponding four-fermion operators 
as $SU(2)_L\otimes SU(2)_R$ (whose subgroup the former is). On the other
hand, the flavon VEV-induced operators can be adequately suppressed by
introducing an additional $U(1)_V$ flavor gauge symmetry (or its appropriate
discrete subgroup). Thus, flavor violation in the TSSM appears to be in the
acceptable range for $M_s\sim 10-100~{\mbox{TeV}}$.

{}One of the many remaining  
open questions is how to explicitly embed the TSSM (or its
variations) in the brane world framework. That is, it would be nice to have
an explicit string construction of such a model. One of the promising
directions in this regard appears to be a possible embedding into the Type
I (Type I$^\prime$) framework. The recent progress in understanding four
dimensional Type I compactifications \cite{typeI,3gen,3gen1} 
raises hope that this might not be out of reach. However, as was pointed
out in \cite{zura,zura1}, if there exists an embedding of the TSSM in the
Type I framework, it appears to be within {\em non-perturbative} Type I
compactifications. A better understanding of such Type I (as well as four
dimensional F-theory \cite{vafa}) compactifications is therefore more than
desirable.

\acknowledgements

{}I would like to thank Nima Arkani-Hamed, Tom Banks and especially Gia
Dvali for useful discussions. 
This work was supported in part by the grant
NSF PHY-96-02074, 
and the DOE 1994 OJI award. I would also like to thank Albert and 
Ribena Yu for financial support.

\end{document}